\documentclass{PoS}
\usepackage{lineno}

\title{Observation of Anisotropy in the Arrival Direction Distribution of TeV Cosmic Rays with HAWC}

\ShortTitle{TeV-Energy Cosmic-Ray Anisotropy with HAWC}

\author{Segev Y. BenZvi,$^a$ \speaker{Daniel W. Fiorino}$^b$
     and Stefan Westerhoff$^b$ for the HAWC Collaboration$^c$\\
     \llap{$^a$}Department of Physics and Astronomy, University of Rochester, Rochester, NY, USA \\
     \llap{$^b$}Department of Physics and WIPAC, University of Wisconsin-Madison, Madison, WI, USA \\
     \llap{$^c$}For a complete author list, see
                \href{http://www.hawc-observatory.org/collaboration/icrc2015.php}{www.hawc-observatory.org/collaboration/icrc2015.php}. \\
     E-mail:  \email{fiorino@wipac.wisc.edu}}

\abstract{The High-Altitude Water Cherenkov (HAWC) Observatory, located 4100 m above sea level
near Sierra Negra (19$^\circ$ N) in Mexico, is sensitive to gamma rays and cosmic rays at TeV
energies. The arrival direction distribution of cosmic rays at these energies shows significant
anisotropy on several angular scales, with a relative intensity ranging between 
$10^{-3}$ and $10^{-4}$. We present the results of a study of cosmic-ray anisotropy based on more 
than 86 billion cosmic-ray
air showers recorded with HAWC since June 2013. The HAWC cosmic-ray sky map, which has a median
energy of 2 TeV, exhibits several regions of significantly enhanced cosmic-ray flux. We present
the energy dependence of the anisotropy and the cosmic-ray spectrum in the regions of significant
excess.}

\FullConference{The 34th International Cosmic Ray Conference,\\
		30 July- 6 August, 2015\\
		The Hague, The Netherlands}

\begin{document}

\section{Introduction}

The HAWC Observatory has recently completed construction 
at 4100 m above sea level near Puebla, Mexico. The observatory, 
located at 19$^\circ$N latitude, is designed to study the sky in gamma rays and
cosmic rays between 50 GeV and 100 TeV.
While cosmic rays are the major source of background
in the gamma-ray analysis, the distribution of the arrival
directions of the cosmic rays is itself of significant interest.
Cosmic rays with rigidities of several TV strongly scatter on 
Galactic magnetic fields ($r_g \approx0.001$~pc), 
scrambling their initial direction many times before reaching Earth.
Despite this, the last decade of experiments
has reported stable $10^{-3}-10^{-4}$ sidereal anisotropy in the arrival direction 
distribution of TeV cosmic rays 
in both hemispheres 
(see Ref~\cite{Abeysekara:2014sna} and references therein).
The anisotropy has been observed at large ($> 60^\circ$) 
and small angular scales by multiple experiments.

At TeV energies, the large-scale structure is dominated by dipole and
quadrupole moments of consistent phase throughout all experiments. These
studies require year-long observations to cancel the effects of the Solar
dipole. This study will not be presented with this half-year data set.
After removal of the large-scale structure, significant small-scale structure
appears, most notably a region of excess near 
($\alpha=60^\circ,\delta=-5^\circ$)~\cite{Abdo:2008aw,ARGO-YBJ:2013gya}.
Small-scale deficits do not appear to be significant features and may be an 
artifact of the removal of the large-scale structure.

The origin of the anisotropy is not well understood. It has 
been suggested that weak dipole or dipole-like features should be a 
consequence of the diffusion of cosmic rays from nearby sources in the
galaxy~\cite{Erlykin:2006ri,Blasi:2011fm}. This hypothesis is currently of 
considerable interest~\cite{Pohl:2012xs,Sveshnikova:2013ui,PhysRevLett.114.021101}.
Excesses may be the result of magnetically connected cosmic-ray
accelerators or something more exotic~\cite{Perez-Garcia:2013lza,Harding:2013qra}. 
Small-scale anisotropy may also arise as a feature of turbulent scattering of 
large-scale anisotropy~\cite{Giacinti:2011mz,Ahlers:2013ima}. 

The newly finished HAWC detector observes a larger portion of the sky with a higher
recorded event rate than any high-energy cosmic-ray detector before it.
This proceeding serves as an update to Ref.~\cite{Abeysekara:2014sna}. 
These data cover a region of the sky previously unobserved
by experiments operating in the northern and southern
hemispheres. In these proceedings we present measurements
of anisotropy on small angular scales with an emphasis on
the energy-dependence of the significant features. 
We compare the observed anisotropy with previous
measurements of the northern and southern skies.

\section{Data Set}

The HAWC Observatory is a 22,000 m$^2$
array of close-packed
water Cherenkov detectors (WCDs). Each WCD
consists of a cylindrical steel water tank 4.5 m in height
and 7.3 m in diameter. 
The complete detector now comprises 300 WCDs with 1200 PMTs.

The analysis presented in this paper uses the data
collected during the operation of 111 WCDs (HAWC-111) 
between June 16, 2013 and July 8, 2014. 
Triggers for gamma-ray and cosmic-ray air showers were
formed with a simple multiplicity trigger which requires >
15 PMTs to be above threshold within a sliding time window
of 150 ns. The trigger rate in HAWC-111 was approximately
15 kHz. 

A subset of the HAWC-111 data set was used for the anisotropy search.
A cut in the fractional number of PMTs > 6\% is used to remove
poorly reconstructed events from the data. Additionally,
only continuous sidereal days of data were chosen for 
in order to reduce the bias of uneven exposure
along right ascension. The resultant data set has a livetime of 
181 days with $8.6 \times 10^{10}$ well-reconstructed
cosmic-ray air showers. Using the detector simulation we estimate that the 
median energy of the data set is about 2 TeV. We estimate the angular 
resolution for the primary particle direction to be approximately 1.0$^\circ$
from both simulation and observation of the cosmic-ray Moon shadow. This is
sufficient to observe the $5^\circ-180^\circ$ features in the 
anisotropy of the cosmic rays.

\section{Analysis}

To search for anisotropy, we directly
compute the relative intensity as a function of equatorial
coordinates ($\alpha$,$\delta$). We begin by binning the sky into an
equal-area grid with a resolution of 0.2$^\circ$ per bin using the 
HEALPix library. A binned 
data map $N(\alpha,\delta)_i$ is used to store the arrival directions of 
air showers recorded by the detector for each angular bin $i$.

To produce residual maps of the anisotropy of the arrival
directions of the cosmic rays, we must have a 
description of the arrival direction distribution if the
data arrived isotropically at Earth, $\langle N \rangle (\alpha,\delta)_i$. 
We calculate this expected
flux from the data themselves in order to account for
all-sky rate variations and the changing viewing angle of each pixel.
The spatial distribution of events is unique to a 
detector configuration and stable for periods exceeding 24 hours.
The all-sky rate varies in Solar time by  $\sim5$\% and approximates
a sinusoid. 

The reference map $\langle N \rangle (\alpha,\delta)_i$ is produced 
using the direct integration
technique described in Ref.~\cite{Atkins:2003ep}, adapted for the HEALPix
grid. In brief, we proceed by collecting all events recorded
during a predefined time period $\Delta t$ and integrate the local
arrival direction distribution against the detector event rate.
The method effectively smooths out the true arrival direction
distribution in right ascension on angular scales of roughly
$\Delta t \cdot 15^\circ$ hr$^{-1}$
such that the analysis is only sensitive to
structures smaller than this characteristic angular scale. The
direct integration procedure also compensates for variations
in the detector rate while preserving the event distribution
in declination.
Once the reference map is obtained, we calculate the
deviations from isotropy by computing the relative intensity

\begin{equation}
\delta I(\alpha,\delta)_i = \frac{\Delta N}{\langle N \rangle}=\frac{N(\alpha,\delta)_i - \langle N \rangle(\alpha,\delta)_i}{\langle N \rangle (\alpha,\delta)_i}~,
\end{equation}
which gives the amplitude of deviations from the isotropic
expectation in each angular bin $i$. The significance of the
deviation can be calculated using the method of Li and
Ma~\cite{Li:1983fv}.

This analysis method can be sensitive to any maximum angular 
scale through the choice of $\Delta t$. Due to the sampling of the
reference map along lines of right ascension, the maximum angle
scales as $(\cos\delta)^{-1}$. Only a choice of 24 hours ensures
a uniform angular scale as a function of declination. To remove larger
structure, a multipole fit can be subtracted to access lower angular 
scales while preserving the maximum angular scale throughout the map.

\section{Measurement of the Small-scale Anisotropy}

\begin{figure}
     \centering
     \includegraphics[width=.6\textwidth]{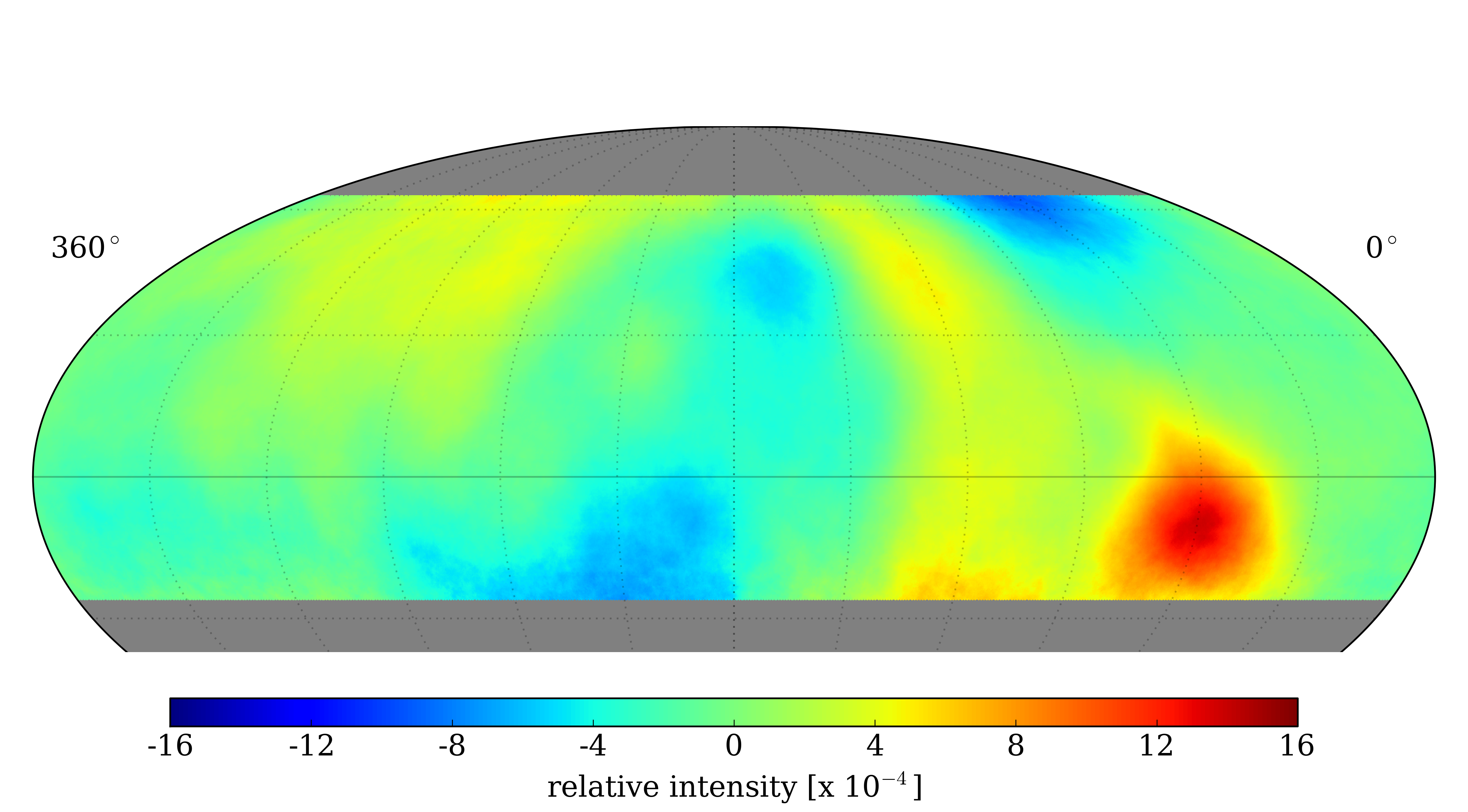}
     \caption{Relative intensity of the cosmic-ray flux for 181 days of HAWC-111, in
equatorial coordinates. A period $\Delta t$ = 24 hr is used to access the largest features present in
the map. The map is shown with 10$^\circ$ smoothing applied.}
     \label{fig:fig1}
\end{figure}

The analysis was carried out on HAWC-111 data using
$\Delta t$= 24 hr to obtain sensitivity to all angular features equally 
over the sky. The results are plotted in Figure~\ref{fig:fig1} in relative 
intensity. The data have been smoothed using a 10$^\circ$ top-hat function to 
make the clustering of arrival directions readily apparent. 
At this time we do not provide an in-depth description of the observed large-scale 
features. The measurement in Fig.~\ref{fig:fig1} is a combination of sidereal 
anisotropy and the Solar dipole effect which causes an excess of cosmic rays in 
the direction of the Earth's motion around the Sun.
The Solar contamination in this data set has been confirmed by observing
significant signal in the unphysical coordinate system using ``anti-sidereal" time.
The two are difficult to disentangle without a full year of data in 
which the signals make a complete transit in the other reference frame.

\begin{figure}
     \centering
     \includegraphics[width=.6\textwidth]{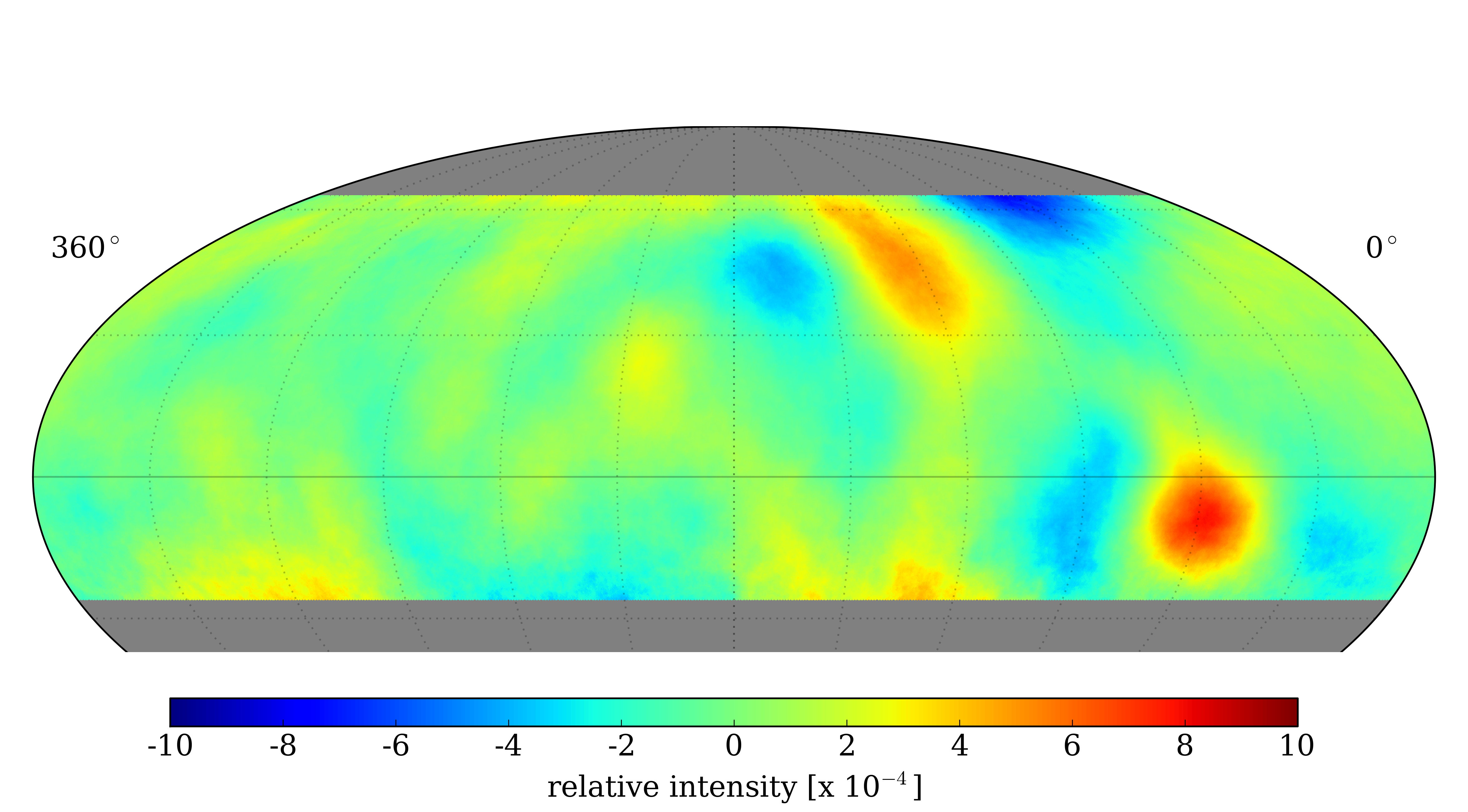}\\
     \includegraphics[width=.6\textwidth]{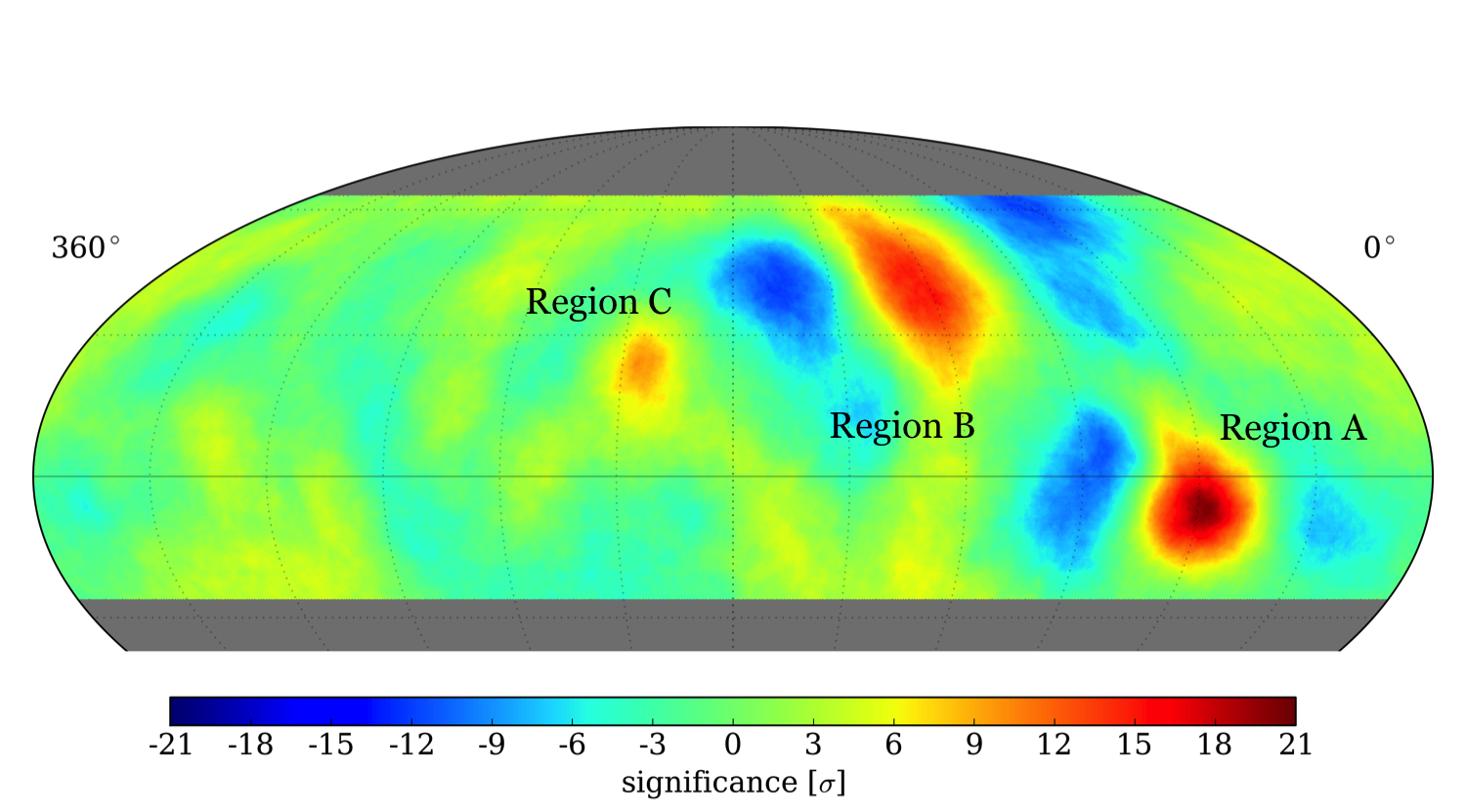}
     \caption{Relative intensity (\emph{top}) and pre-trial significance 
(\emph{bottom}) of the cosmic-ray flux
after fit and subtraction of the dipole, quadrupole, and octupole terms from the map shown
in Fig.~1. The map is shown with $10^\circ$ smoothing applied.}
     \label{fig:fig2}
\end{figure}

\begin{figure}
     \centering
     \includegraphics[width=.6\textwidth]{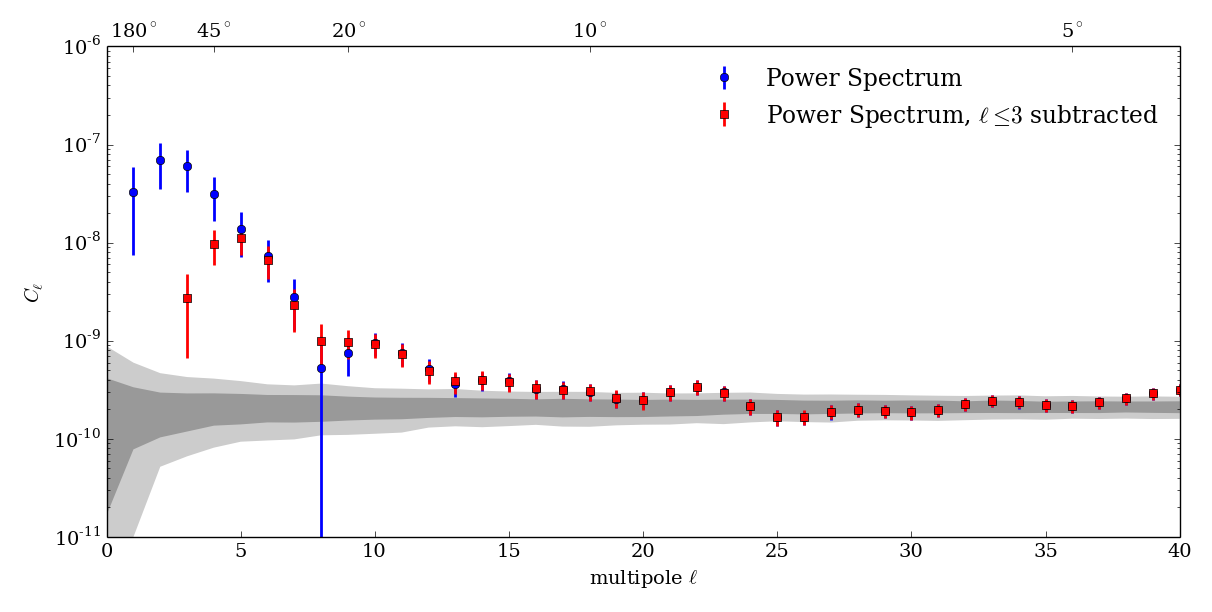}\\
     \caption{Angular power spectra of the unsmoothed relative intensity map before
(blue) and after (red) subtracting the large-scale structure ($\ell \leq 3$). 
Gray errors bands show the 68\% and 95\% spread of the $C_\ell$ for isotropic data sets.
Comparing the band to the data shows which $ell$-modes significantly contribute to the sky map.
The error bars on the $C_\ell$ are the square root of the variances returned by a fit using
a power spectrum estimator (PolSpice).}
     \label{fig:fig3}
\end{figure}

The large-scale signals can be subtracted from
the map in Fig.~\ref{fig:fig1} using a multipole fit to all multipole moments 
$\ell \leq \ell_{max}$. Any small-scale residual 
from the Solar signal should be negligible ($< 10^{-5}$), so we are left with the 
small-scale structure in the sidereal frame. The residual structure will contain 
power at angular scales less than the $180^\circ/\ell_{max}$. 
To reveal angular scale $<60^\circ$, a choice of $\ell_{max}=3$ was made.
The fit of all moments with $\ell \leq 3$ in the map in Fig.~\ref{fig:fig1} were
was subtracted to produce the relative intensity and significance maps shown in 
Figure~\ref{fig:fig2}. Again, a $10^\circ$ smoothing 
has been applied, so only pre-trials significances are shown. 
The estimated number of trials is at most equal to the number of independent bins ($\sim10^5$),
but is in fact much smaller, since we are not performing a blind search; all excess regions have
been perviously observed by other experiments.
The subtraction of the fit removes angular
power in $\ell \leq 3$  without influencing the higher order multipoles as 
evidenced by a comparison of the angular power spectra in Figure~\ref{fig:fig3}. 

With the trials factor taken into account, three significant features 
remain, the strong region of excess flux at $(\alpha = 60^\circ,\delta=-10^\circ)$, the band of
excess along $\alpha = 120^\circ$, and a weaker excess at $(\alpha=240^\circ,\delta=45^\circ)$.
These excesses lie coincident with the $10^\circ-20^\circ$ regions of cosmic-ray access
first observed by Milagro (Regions A and B in \cite{Abdo:2008aw}) and
a third region observed only by ARGO-YBJ~\cite{ARGO-YBJ:2013gya} (Region C).
The strongest excess, with a pre-trial significance of 
$20.6 \sigma$, is found at $\alpha = 58.4^\circ$ and $\delta = -8.8^\circ$ 
and corresponds to Region A in the Milagro map. The relative intensity of the excess 
in this region peaks at 
$(7.9 \pm 0.4 \pm 0.4)\times 10^{-4}$,
where the first error is statistical and the second error is systematic. 
The median cosmic-ray energy at this declination is 2.1 TeV. For comparison, 
we fit a 2-D Gaussian function to the relative intensity map around Region A. 
The center is located at $\alpha = 59.2^\circ \pm 0.6^\circ$ and 
$\delta = -7.2^\circ \pm 0.7^\circ$, with an amplitude of 
$(8.7 \pm 0.9) \times 10^{-4}$. The width is  nearly symmetric in $\alpha$
and $\delta$, approximately $7.6^\circ \pm 1.2^\circ$. 

The elongated excess around $\alpha = 120^\circ$, coincident with Region B, 
extends over a wide range of declinations. It is most significant 
at $(119.5^\circ,38.9^\circ)$ with a 
pre-trial significance of 15.6$\sigma$ and a relative intensity of 
$(5.3 \pm 0.4 \pm 0.2) \times 10^{-4}$.
A third excess region, Region C, is centered at 
$\alpha = 206.0^\circ$ and $\delta = 23.8^\circ$ with a
pre-trial significance of 10.5$\sigma$ and a peak relative 
intensity of $(2.8 \pm 0.3 \pm 0.7) \times 10^{-4}$. 
The median cosmic-ray energy at this declination is 2.0 TeV.

There are also regions of strong deficits visible, typically on both 
sides of the strong excess regions. The appearance of these deficit 
regions, correlated with the excess regions, is likely an
artifact caused by imperfections in the multipole fit.

\section{Study of Region A}

An early version of energy estimation has been developed for studying the
HAWC data set as a function of energy. Seven energy bins are separated 
by cutting on both the incident zenith angle and the fraction of the detector that
observed a cosmic-ray air shower. 
The energies of these bins are determined from
simulation. At this time, the energy bins are
heavily overlapping, still, a clear evolution of
Region A as a function of energy is identifiable. This region exhibits a harder
spectrum than background.

\begin{figure}
     \centering
     \includegraphics[width=.6\textwidth]{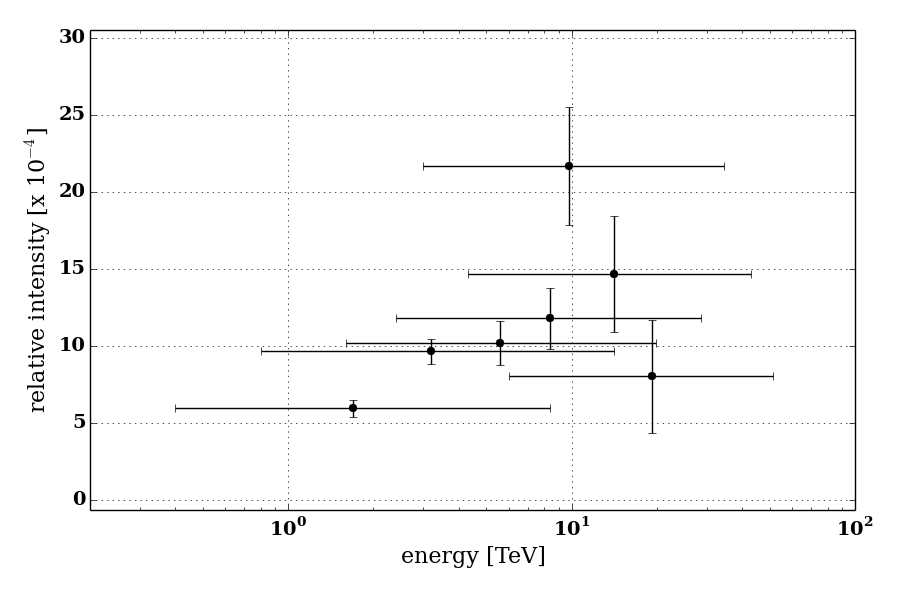}\\
     \caption{Spectrum of Region A in relative intensity in different energy-proxy bins. The
energies of the data were determined from simulation. The error bars on the median energy
values correspond to a 68\% containing interval.}
     \label{fig:fig4}
\end{figure}

We estimate the statistical significance of the hardness the in Region 
A spectrum by comparing the slope of a linear fit of $\delta I$ versus $\log(E)$ 
in Figure~\ref{fig:fig4} to similar fits performed at many random
locations in the field of view. The random locations excluded a 15$^\circ$ 
circle centered on Regions A, B, and C. The distribution of slopes for 
the random locations follows a Gaussian
distribution centered at zero with a width of $0.9 \times 10^{-4}$. The 
best fit slope at the position of Region A is $(3.8 \pm 1.1) \times 10^{-4}$, 
4.2$\sigma$ away from the mean. The reduced $\chi^2$ of a fit to a sloped-line 
is 1.16 compared to 5.66 for a horizontal line. This further supports a spectrum 
that is harder than background. 

\begin{figure}
  \centering
  \includegraphics[width=\textwidth]{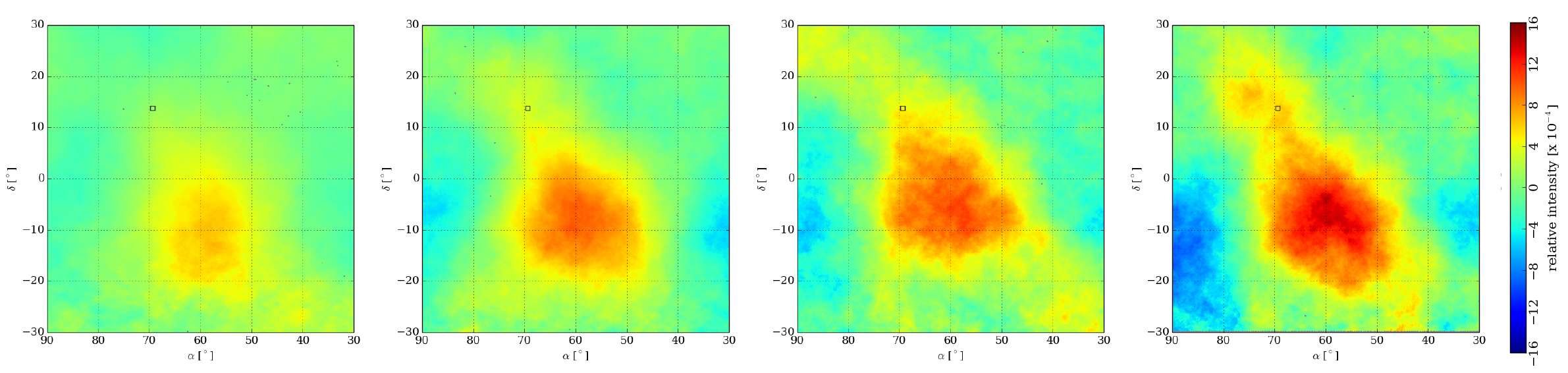}
  \caption[Region A in four energy-proxy bins]
  {Relative intensity of Region A for 4 different energy-proxy bins. The square mark
  denotes the location of the centroid of Region A as reported by Milagro ($\alpha=69.4^\circ$,
  $\delta=13.8^\circ$) at 10 TeV.  The median energy of the data in each plot from left to right is
  ${1.7}^{+6.6}_{-1.3}$\,TeV, 
  ${3.2}^{+10.9}_{-2.4}$\,TeV,
  ${5.6}^{+14.2}_{-3.9}$\,TeV, and
  ${14.1}^{+28.7}_{-9.9}$\,TeV.
  }
  \label{fig:fig5}
\end{figure}

A zoomed view of Region A in four energy-proxy 
bins (the final four bins of Fig.~\ref{fig:fig4} were combined) is provided 
in Figure~\ref{fig:fig5}. In addition to the spectral steepening, this figure
shows the excess extending to higher declination as a function of energy,
consistent with the ARGO observation.

\section{Discussion}

This half-year HAWC-111 data set is already one third as large as the 
Milagro 8-year data set and one fourth of the ARGO-YBJ 5-year data set. 
We have found significant small-scale
structure that coincides with 3 of the 4 previously published regions of 
cosmic-ray excess. In general, there is excellent agreement between 
the observed excesses and deficits of previous measurements by ARGO-YBJ
and Milagro in the Northern Hemisphere. 

Region A at $\alpha=60^\circ$ appears to have two sub-regions in
the TeV range. At higher energies (>10 TeV),
the northern part at $\delta=5^\circ$ appears. 
Below this, the southern part at $\delta=-5^\circ$ is dominant. Milagro
saw the northern part of Region A, but only observed down to $\delta=-5^\circ$.
There is a slight indication of the southern part of Region A in the unpublished
relative intensity maps. 
Both the HAWC and ARGO maps see the change with energy in Region A. 
This suggests that the hardening of the spectrum of Region A is coincident 
with a northernly movement of the excess or the emergence of an entirely new
region of excess. 

Region B at $\alpha$ = 120$^\circ$ is visible at all declinations
observed by HAWC and seems to follow the edge of the large-scale sidereal
deficit. The extension of the excess to higher
declinations is notable because this is not observed
to high significance
in the Milagro data, even though Milagro was located at
a higher latitude ($35^\circ$N). The significance maps published
by ARGO also extend to high declinations, and these
maps indicate that the Region B excess extends northward.

Region C at $\alpha$ = 205$^\circ$ is not significant in the Milagro data, 
but the ARGO collaboration has observed a hot spot at the same location of 
similar shape and intensity.
ARGO also observed a fourth significant 
region at $\alpha=280^\circ$ which appears as a sub threshold excess in the
HAWC map. Continued observation with HAWC will help confirm this as 
a true excess.

A comparison between data from HAWC-111 and IceCube (20 TeV median)
in the Southern Hemisphere is very interesting because
connecting the northern and southern measurements would eliminate biases 
from partial sky coverage. No clear connection of the
small-scale anisotropy that is present in both hemispheres has 
been made yet. The published IceCube maps are of higher energy than 
HAWC and other northern measurements. Because of this, a first effort 
at combining HAWC and IceCube data
uses cuts to bring their median energies closer to a 
central value~\cite{DiazVelez:2015}.

Ten years after the first sky maps that
charted the cosmic-ray anisotropy, the state of the art in this 
field has moved towards studying significant cosmic-ray features 
as a function of rigidity.
Precision measurements of cosmic-ray anisotropy are readily obtained
from air shower detectors, such as HAWC. Initial improvements to the
resolution and range of measured cosmic-ray energies using these detectors 
have revealed energy-dependent behavior of the anisotropy. Further studies of 
this phenomenon may provide insight into local cosmic-ray production
and propagation or may even become a tool for studying exotic physics.

\bibliography{icrc2015-0147}

\providecommand{\href}[2]{#2}\begingroup\raggedright\begin{thebibliography}{10}

\bibitem{Abeysekara:2014sna}
{\bf HAWC} Collaboration, A.~Abeysekara et~al., {\it {Observation of
  Small-scale Anisotropy in the Arrival Direction Distribution of TeV Cosmic
  Rays with HAWC}},  {\em The Astrophysical Journal} {\bf 796} (2014) 108,
  [\href{http://arxiv.org/abs/1408.4805}{{\tt arXiv:1408.4805}}].

\bibitem{Abdo:2008aw}
{\bf Milagro} Collaboration, A.~A. Abdo et~al., {\it {The Large Scale
  Cosmic-Ray Anisotropy as Observed with Milagro}},  {\em Astrophys. J.} {\bf
  698} (2009) 2121--2130, [\href{http://arxiv.org/abs/0806.2293}{{\tt
  arXiv:0806.2293}}].

\bibitem{ARGO-YBJ:2013gya}
{\bf ARGO-YBJ} Collaboration, B.~Bartoli et~al., {\it {Medium scale anisotropy
  in the TeV cosmic ray flux observed by ARGO-YBJ}},  {\em Phys.Rev.} {\bf D88}
  (2013) 082001, [\href{http://arxiv.org/abs/1309.6182}{{\tt
  arXiv:1309.6182}}].

\bibitem{Erlykin:2006ri}
A.~D. Erlykin and A.~W. Wolfendale, {\it {The anisotropy of galactic cosmic
  rays as a product of stochastic supernova explosions}},  {\em Astropart.
  Phys.} {\bf 25} (2006) 183--194,
  [\href{http://arxiv.org/abs/astro-ph/0601290}{{\tt astro-ph/0601290}}].

\bibitem{Blasi:2011fm}
P.~Blasi and E.~Amato, {\it {Diffusive propagation of cosmic rays from
  supernova remnants in the Galaxy. II: anisotropy}},  {\em JCAP} {\bf 1201}
  (2012) 011, [\href{http://arxiv.org/abs/1105.4529}{{\tt arXiv:1105.4529}}].

\bibitem{Pohl:2012xs}
M.~Pohl and D.~Eichler, {\it {Understanding TeV-Band Cosmic-Ray Anisotropy}},
  {\em Astrophys. J.} {\bf 766} (2013) 4,
  [\href{http://arxiv.org/abs/1208.5338}{{\tt arXiv:1208.5338}}].

\bibitem{Sveshnikova:2013ui}
L.~G. Sveshnikova, O.~N. Strelnikova, and V.~S. Ptuskin, {\it {Spectrum and
  Anisotropy of Cosmic Rays at TeV-PeV-energies and Contribution of Nearby
  Sources}},  {\em Astropart. Phys.} {\bf 50-52} (2013) 33--46,
  [\href{http://arxiv.org/abs/1301.2028}{{\tt arXiv:1301.2028}}].

\bibitem{PhysRevLett.114.021101}
P.~Mertsch and S.~Funk, {\it {Solution to the Cosmic Ray Anisotropy Problem}},
  {\em Phys. Rev. Lett.} {\bf 114} (Jan, 2015) 021101.

\bibitem{Perez-Garcia:2013lza}
M.~{\'A}ngeles P{\'e}rez-Garc{\'i}a, K.~Kotera, and J.~Silk, {\it {Anisotropy
  in Cosmic rays from internal transitions in neutron stars}},  {\em Nucl.
  Instrum. Meth.} {\bf A742} (2014) 237--240,
  [\href{http://arxiv.org/abs/1309.1852}{{\tt arXiv:1309.1852}}].

\bibitem{Harding:2013qra}
J.~P. Harding, {\it {The TeV Cosmic-Ray Anisotropy from Local Dark Matter
  Annihilation}},  {\em arXiv:1307.6537} (2013)
  [\href{http://arxiv.org/abs/1307.6537}{{\tt arXiv:1307.6537}}].

\bibitem{Giacinti:2011mz}
G.~Giacinti and G.~Sigl, {\it {Local Magnetic Turbulence and TeV-PeV Cosmic Ray
  Anisotropies}},  {\em Phys. Rev. Lett.} {\bf 109} (2012) 071101,
  [\href{http://arxiv.org/abs/1111.2536}{{\tt arXiv:1111.2536}}].

\bibitem{Ahlers:2013ima}
M.~Ahlers, {\it {Anomalous Anisotropies of Cosmic Rays from Turbulent Magnetic
  Fields}},  {\em Phys. Rev. Lett.} {\bf 112} (2014) 021101,
  [\href{http://arxiv.org/abs/1310.5712}{{\tt arXiv:1310.5712}}].

\bibitem{Atkins:2003ep}
{\bf Milagro} Collaboration, R.~W. Atkins et~al., {\it {Observation of TeV
  gamma-rays from the Crab nebula with MILAGRO using a new background rejection
  technique}},  {\em Astrophys. J.} {\bf 595} (2003) 803--811,
  [\href{http://arxiv.org/abs/astro-ph/0305308}{{\tt astro-ph/0305308}}].

\bibitem{Li:1983fv}
T.-P. Li and Y.-Q. Ma, {\it {Analysis methods for results in gamma-ray
  astronomy}},  {\em Astrophys. J.} {\bf 272} (1983) 317--324.

\bibitem{DiazVelez:2015}
{\bf HAWC} Collaboration, J.~D{\`i}az-Velez and D.~Fiorino, {\it {Full-Sky
  Analysis of Cosmic-Ray Anisotropy with IceCube and HAWC}},  in {\em Proc.
  34th ICRC}, (The Hague, The Netherlands), August, 2015.

\end{thebibliography}\endgroup

\end{document}